\documentclass[iop]{emulateapj}
\usepackage{graphicx,epsfig,epsf}
\usepackage{graphicx}
\usepackage{times}
\usepackage{url}

\makeatletter
\def\url@leostyle{%
\@ifundefined{selectfont}{\def\UrlFont{\sf}}{\def\UrlFont{\small\ttfamily}}}
\makeatother
\bibpunct[;]{(}{)}{;}{a}{}{,}

\def\p  {\phantom{0}}

\def\NM   {NovaMus}

\def\Vlsr {\ifmmode {V_{\rm LSR}} \else {$V_{\rm LSR}$} \fi}
\def\Ro   {\ifmmode {R_0} \else {$R_0$} \fi}
\def\To   {\ifmmode {\Theta_0} \else {$\Theta_0$} \fi}

\def\p    {\phantom{0}}
\def\simless{\lower2pt\hbox{$\buildrel {\scriptstyle <}
   \over {\scriptstyle\sim}$}}
\def\pd  {\lower0pt\hbox{$\buildrel {^\circ} \over {.}$}}

\def\aap{A\&A}

\def\aj{AJ}
\def\apj{ApJ}
\def\apjl{ApJ}
\def\apjs{ApJS}

\def\araa{ARA\&A}

\def\mnras{MNRAS}

\def\pasj{PASJ}
\def\pasp{PASP}

\newcounter{lastnote}

\begin{document}
\title{\sc The Spin of The Black Hole in the X-ray Binary Nova Muscae 1991}

\shorttitle{The Spin of Nova Muscae 1991}
\shortauthors{Chen et al.}

\author{Zihan~Chen,$^{1,2}$  Lijun~Gou,$^{1,2}$ Jeffrey E. McClintock,$^{3}$
  James F. Steiner,$^{3}$ Jianfeng~Wu,$^{3}$ Weiwei~Xu,$^{1,2}$ \\
  Jerome A. Orosz,$^{4}$ Yanmei Xiang$^{1,5}$}
\altaffiltext{1}{National Astronomical Observatories, Chinese Academy of Sciences, 
				 Beijing 100012, China}
\altaffiltext{2}{University of Chinese Academy of Sciences, Beijing
  100012, China}
\altaffiltext{3}{Harvard-Smithsonian Center for Astrophysics, 60 Garden Street, Cambridge, MA 02138, USA}
\altaffiltext{4}{Department of Astronomy, San Diego State University, 5500 Campanile Drive, San Diego, CA 92182, USA}
\altaffiltext{5}{Department of Astronomy, Key Laboratory of Astroparticle Physics of Yunnan Province, Yunnan University, Kunming, 650091, China}
\slugcomment{ApJ}
\begin{abstract}

  The bright soft X-ray transient Nova Muscae 1991 was intensively
  observed during its entire 8-month outburst using the Large Area
  Counter (LAC) onboard the {\it Ginga} satellite.  Recently, we
  obtained accurate estimates of the mass of the black hole primary, the
  orbital inclination angle of the system, and the distance.  Using
  these crucial input data and {\it Ginga} X-ray spectra, we have
  measured the spin of the black hole using the continuum-fitting
  method. For four X-ray spectra of extraordinary quality we have
  determined the dimensionless spin parameter of the black hole to be
  $a_* = 0.63_{-0.19}^{+0.16}$ (1$\sigma$ confidence level), a result
  that we confirm using eleven additional spectra of lower quality. Our
  spin estimate challenges two published results: It is somewhat higher
  than the value predicted by a proposed relationship between jet power
  and spin; and we find that the spin of the black hole is decidedly
  prograde, not retrograde as has been claimed.

\end{abstract}


\section{Introduction} \label{sect:intro}

On 1991 January 8, a bright X-ray nova was discovered independently
using the {\it Ginga} and {\it Granat} X-ray satellites by
\citet{Kitamoto:92} and \citet{Brandt:92} who named the source GS
1124--683 and GRS 1124--68, respectively. Located in the constellation
Musca, the X-ray source is also known as X-ray Nova Muscae 1991
(hereafter, \NM). After the system returned to quiescence, optical
observations revealed an orbital period of 10.4 hr and a large mass
function \citep{Remillard:92}, which established that the system is one
of about a dozen short-period X-ray binaries ($P_{\rm orb} <
12$~hr) whose compact X-ray source is a dynamically-confirmed black
hole.  The prototype of this subclass of black hole binaries is
A0620--00. Other well studied short-period systems include GRO J0422+32,
XTE J1118+480, XTE J1859+228 and GS 2000+25.

During the past decade, the spins of many black holes have been
estimated using two methods: fitting the profile of the Fe K line
\citep{Fabian:89,Reynolds:14} and fitting the thermal continuum spectrum
\citep{Zhang:97,McClintock:14}{\footnote{Spin is usually expressed in
terms of the dimensionless black-hole spin parameter $a_* = cJ/GM^2$,
where $a_*$ is subject to the Kerr bound $|a_*|<1$, and $J$ and $M$ are
respectively the angular momentum and mass of the black hole.}.  It is
the continuum-fitting method that we employ here in measuring the spin
of \NM, and that our group has developed and used to measure the spins
of ten stellar-mass black holes \citep{McClintock:14,Steiner:14,Gou:14}.

In the continuum-fitting method, the spin of a black hole with known
mass and distance is estimated by fitting the thermal component of
emission to a corrected version of the thin-disk model of Novikov and
Thorne \citep{Li:05} while employing an advanced treatment of spectral
hardening \citep{Davis:05,Davis:06}. For the successful application of
the method, it is essential to consider only those spectra that contain
a dominant thermal component \citep{Steiner:09a} and for which the
Eddington-scaled disk luminosity is moderate, $l \equiv L_{\rm
bol}(a_{\ast}, \dot{M})/L_{\rm Edd}< 0.3$ \citep{McClintock:06}.  

The robustness of the continuum-fitting method has been demonstrated by
the very many independent and consistent measurements of spin that have
been obtained for several black holes \citep[e.g.,][]{Steiner:10}; by
extensive theoretical studies of the thin-disk model
\citep{Shafee:08,Penna:10,Noble:11,Kulkarni:11,Zhu:12}; and through
careful consideration of a wide range of systematic errors \citep[~and
references therein]{McClintock:14}. In applying the method, one must
usually make the weakly-tested assumption that the spin of the black
hole is closely aligned with the angular momentum vector of the inner
disk \citep{Steiner:12,Fragos:10}.

A crucial requirement of the continuum-fitting method is that one have
accurate estimates of three system parameters: the black hole mass $M$,
the disk inclination $i$ and the source distance $D$. Using optical
dynamical data of unprecedented quality and published light curves, we
have obtained definitive measurements of these parameters for \NM: $
M=11.0^{+2.1}_{-1.4}\p M_\sun$,\p$\rm i=43.2^{+2.1}_{-2.7}\p\rm deg$,\p$
D=4.95^{+0.69}_{-0.65}\p\rm kpc$ \citep{Wu:15b}.  With the values of
these key parameters in hand, in the following sections we fit X-ray
data obtained during the 1991 outburst of \NM, estimate the spin of its
black hole, and present and discuss our results.

We earlier estimated the spin of the black hole in the prototype of the
short-period systems, A0620--00, and found it to be exceptionally low
$a_*=0.12 \pm 0.19$ \citep{Gou:10}.  Another notable feature of this
study of A0620--00 was the extreme dominance of the thermal disk
component: The Compton component contributed $<1$\% of the total flux, a
situation that contrasts sharply, for example, with the case of Cyg X-1
where the Compton component is always unfavorably high
\citep[$>10$\%;][]{Gou:11,Gou:14}.  For \NM, we analyze 15 spectra, four
of which we refer to as our ``gold'' spectra because their Compton
component is minuscule, contributing $<0.05$\% of the total emission. In
this circumstance, how one chooses to model the nonthermal component of
emission is completely irrelevant, the analysis is simple and the
results are particularly robust. Our estimate of spin for \NM\ is based
entirely on our analysis of these four gold spectra, a result we confirm
by analyzing the remaining eleven ``silver'' spectra whose Compton
component is in the range 0.3\% to 9\%.

  The paper is organized as follows. We discuss our data selection
  and their reduction in Section~\ref{sect:data}. In
  Section~\ref{sect:analysis} we describe our methods of data analysis
  and present our results, first for the four gold spectra, and then for
  the complete sample of 15
  spectra. Sections~\ref{sect:error}~and~\ref{sect:error15} are devoted
  respectively to a comprehensive error analysis of the four gold
  spectra, and then to the complete sample of 15 spectra. In
  Section~\ref{sect:discussion} we discuss and summarize our results.

\section{Data Selection \& Reduction} \label{sect:data}

The X-ray data we consider for NovaMus are those presented in
\citet[~hereafter EB94]{Ebisawa:94}, which were obtained using the Large
Area Counter (LAC) onboard the Japanese X-ray astronomy satellite {\it
  Ginga}.  The LAC was comprised of eight identical proportional counter
detectors with a total effective area of 4000 $\rm cm^2$ covering the
energy range from 2.0 keV to 37 keV.

We have strictly followed the procedures described in the manual {\it
  ABC Guide to the Ginga Data
  Analysis}\footnote{www.darts.isas.jaxa.jp/astro/ginga/analysis.html}.
The data were first cleaned and then the spectra were extracted using
the software package {\sc ISAS}. Among the three methods of extracting
the spectra -- the Simple Method, the SUD-sort Method and the Hayashida
Method -- we chose the latter mainly because the approach is
straightforward.  \citet{Hayashida:89} developed an accurate model of
the background that reproduces the background rate for each energy
channel, thereby making it unnecessary to extract separate background
spectra.  This approach is particularly well-suited in the case of a
bright source like \NM, which reached a peak 1--6 keV X-ray intensity of
$\sim8$~Crab \citep{Kitamoto:92}.  The extracted spectra were saved in
ASCII format and (in order to be compatible with XSPEC) were converted
to FITS format using the ftool {\it lac2xspec}, which was also used to
calculate the relevant response files.  As customary, we added a 2\%
systematic error in each channel in quadrature with the statistical
error.  In addition, we binned the spectra to contain at least 25 counts
per channel to insure the validity of the $\chi^2$ statistic.

Because the continuum-fitting method relies on an accurate estimate of
luminosity, we corrected the effective area of the LAC using the
spectrum of the Crab Nebula as a standard source by the method described
in \citet{Steiner:10}.  Specifically, we analyzed one LAC spectrum of
the Crab obtained near the time of the observations in question;
compared the fit parameters obtained to those of our reference Crab
spectrum of \citet[][~$\Gamma = 2.1$ and $N = 9.7 \p\rm photons\p
s^{-1}~cm^{-2}~keV^{-1}$]{Toor:74}; and computed a pair of correction
factors: a normalization correction, $C_{\rm TS}$ = 1.164$\pm$0.024 (the
ratio of the fitted normalization to that of Toor \& Seward) and a
correction to the slope of the power-law, $\Delta \Gamma_{TS}$ =
0.022$\pm$0.009 (the difference between the observed value of the
power-law index and that of Toor \& Seward).  These corrections were
applied in all of our analysis work to each spectrum using the
customized XSPEC multiplicative model {\sc crabcor}.

\section{Data Analysis and Results } \label{sect:analysis}

Twenty-one spectra were extracted initially (see
Table~\ref{table:time}).  However, six were rejected: three (Nos. 5, 15
and 16) because of their short exposure times ($<100$~s) and three
others (Nos. 17, 19 and 20) because their Eddington-scaled luminosities
are $<2$\%.  Table~\ref{table:observations} lists our final sample of 15
time-ordered spectra with the four gold spectra (SP1 -- SP4) listed
first, followed by the 11 silver spectra (SP5 -- SP15).  The errors
considered in this section are those due solely to counting
statistics. The dominant errors due to the uncertainties in the input
parameters $M$, $i$, $D$ and $N_{\rm H}$ are considered in the following
two sections.

All data analysis and model fits were performed using XSPEC version
12.8.2 \citep{Arnaud:96}.  The spectra were fitted over the energy range
2.0~--~25.0 keV.  Because of the detector's limited low-energy response,
we were unable to fit for the hydrogen column density $N_{\rm H}$.  We
estimate this parameter using two published measurements of reddening,
which are consistent within 0.25$\sigma$: E(B-V)~=~$0.287~\pm~0.004$
\citep{Cheng:92} and $0.30~\pm~0.05$ \citep{Shrader:93}.  Adopting this
latter value and its uncertainty, and assuming $A_{\rm V}/E(B-V)~=~3.1$
and $N_{\rm H}/A_{\rm
  V}~=~(2.21~\pm~0.09)\times10^{21}~$mag$^{-1}$cm$^{-2}$ \citep{Guver:09},
we estimate the column density to be $N_{\rm
  H}\rm=(0.206~\pm~{0.035})\times10^{22}~cm^{-2}$.  Throughout the
paper, we use this value of $N_{\rm H}$ and the photoelectric absorption
model {\sc tbabs} with the abundances set using the command {\it abund wilm}
\citep{Wilms:00}.

As in our earlier work on black hole spin
\citep[e.g.,][]{Gou:09,Gou:10}, before embarking on the relativistic
analysis we performed a preliminary nonrelativistic analysis of the
spectra as a check that we have extracted them properly.  To this end,
for \NM\ we compare our results to those of EB94.  Our model is {\sc
  tbabs$\ast$crabcor$\ast$(simpl$\otimes$diskbb)}.  The convolution
model {\sc simpl} is an empirical model of Comptonization with two fit
parameters, the familiar photon index $\Gamma$ and the scattering
fraction -- the fraction $f_{\rm SC}$ of the seed photons that are
scattered into the power-law tail -- which is a straightforward measure
of the strength of the Compton component \citep{Steiner:09b}.  We use
the same nonrelativistic thin-disk model as EB94, namely {\sc diskbb}.
In Table~\ref{table:temperature}, we give for each of the 15 spectra our
estimate of the inner disk temperature $T_1$ and compare our values to
those obtained by EB94 (denoted in the table as $T_2$).  Our results are
consistent with those of EB94 to within about 4\%. It is likely that the
small differences in temperature result from using different models for
the Compton component: We used {\sc simpl} and EB94 used {\sc
  powerlaw}. Support for this view is provided by the four gold spectra
(SP1 -- SP4), which are essentially uncontaminated by power-law
emission; for these spectra, the temperature differences are $<1$\%.

Now, we turn to the relativistic analysis of our four gold
spectra. Replacing {\sc diskbb} by our workhorse relativistic disk model
{\sc kerrbb{\small 2}}, our complete model becomes: \\

{\sc tbabs$\ast$crabcor$\ast$(simpl$\otimes$kerrbb{\small 2})}\\

The components {\sc tbabs}, {\sc crabcor} and {\sc simpl} are described
above.  The key component of the model is {\sc kerrbb{\small 2}}, a thin
accretion-disk model that includes all relativistic effects,
self-irradiation of the disk, limb darkening and the effects of spectral
hardening \citep{Li:05,McClintock:14}.  The two fit parameters of {\sc
  kerrbb{\small 2}} are the spin parameter $a_*$ and the mass accretion
rate $\dot M$.  The effect of spectral hardening is incorporated into
the parent model {\sc kerrbb} via a pair of look-up tables for the
hardening factor $f$ corresponding to two representative values of the
viscosity parameter: $\alpha=0.01$ and 0.1.  The entries in the table
were computed using a second relativistic disk model {\sc bhspec}
\citep{Davis:05,Davis:06}.  In fitting, we turned on the effects of
self-irradiation of the disk (rflag=1) and limb darkening (lflag=1).

In this section, we fix the three external input parameters at their
best-fit values: $M=11.0~M_\sun$, $i=43.2 \rm ~deg$ and $D=4.95\rm ~kpc$
\citep{Wu:15b}.  An inspection of Table~\ref{table:observations} shows
that the fits to the four gold spectra are good, the scattering fraction
negligible, the luminosity $\approx10$\% of the Eddington limit, and
{\it the spin parameter is very precisely determined and lies in a
  narrow range, $a_*=0.61-0.64$, which is the principal result of this
  section.} The much larger error in the spin parameter due to
uncertainties in the parameters $M$, $i$, $D$ and $N_{\rm H}$ is
considered in the following section.

We now confirm our estimate of $a_*$ by presenting results for the
silver spectra (SP5 -- SP15).  Although these spectra are strongly
dominated by the thermal component and quite suitable for application of
the continuum-fitting method, obtaining good fits requires that we add a
minor reflection component.  The complete model we employ is:\\
 	
{\sc tbabs$\ast$crabcor$\ast$(simplr$\otimes$kerrbb{\small 2}\\
	+kerrconv$\otimes$(ireflect$\otimes$simplc)+kerrdisk)}.\\
	
While more complex, the model is similar to the one used in analyzing
the gold spectra. The thin-disk model {\sc kerrbb{\small 2}} is again
decidedly the main component.  The two multiplicative models out front
are the same as before.  Likewise, the first term in parenthesis is the
same except that {\sc simpl} has been replaced by {\sc simplr}, a
variant of {\sc simpl} that computes the Compton component to
accommodate a nonzero reflection fraction, and that has the ability to
isolate the Compton component \citep{Steiner:11}.

The second and third additive terms in parenthesis model the reprocessed
emission from the disk that results from its illumination by the
power-law component. The model for the illuminating power-law component
itself (the term on the far right) is {\sc simplc}, which is equivalent
to {\sc simplr$\otimes$kerrbb{\small 2} } minus the unscattered thermal
component. The component {\sc ireflect} acts solely on the power-law
component to generate the reflection continuum with absorption edges;
key parameters of {\sc ireflect} are the disk ionization parameter $\xi$
and the disk temperature $T$ (which we fix to the value $T_1$ in
Table~\ref{table:temperature}).  To complete the model of the reflected
component, we follow \citep{Brenneman:06} and employ the line model {\sc
  kerrdisk} and the convolution smearing model {\sc kerrconv}, both of
which treat $a_*$ as a free fit parameter. These models allow the
emissivity indices to differ in the inner and outer regions of the
disk. For simplicity, and because this parameter is unknown with values
that vary widely from application to application, we use an unbroken
emissivity profile with a single index q. We tie together all the common
parameters of {\sc kerrdisk} and {\sc kerrconv}, including the two
principal parameters, namely, $a_*$ and $q$. The key parameters of {\sc
  kerrdisk} are the rest-frame line energy $E_{\rm L}$ and the photon
flux in the line $N_{\rm L}$.  At the level of detail, we set the
inclination to our measured value of 43.2 deg; fixed the abundances of
all the elements to solar; set the scaling factor in {\sc simplr} to -1;
allowed the ionization parameter to vary; set the emissivity index to 3
in {\sc kerrconv}; and for {\sc kerrdisk} adopted an unbroken emmisivity
profile with a single index $q$ while linking the spin parameter of this
component to that in {\sc {kerrbb{\small 2}}}. The fitting results to the reflection
 components for SP5 -- SP15 are listed separately in Table~\ref{table:observations_2}.

An inspection of Table~\ref{table:observations} shows that the spin
parameter for the eleven silver spectra is quite precisely determined.
However, its value ranges rather widely from 0.47 to 0.72, while its
average value is $a_* = 0.58 \pm 0.02$ (std.\ dev.), which is in good
agreement with the value found for the gold spectra. The luminosity for
the silver spectra likewise ranges widely, from 3\% to 20\% of the
Eddington limit.

The results given in Table~\ref{table:observations} for all 15 spectra
are for our baseline value of the viscosity parameter, $\alpha=0.1$.  For
our other fiducial value, $\alpha=0.01$, the spin increases slightly and
systematically, as we show in Section~\ref{sect:error}, and as we have
found to be consistently the case in measuring the spins of other black
holes. We assume throughout the paper that the metallicity of the disk
gas is solar, although this parameter has a negligible effect on our
results \citep[e.g., see][]{Gou:10}.

One of the parameters of the reflection model {\sc ireflect} is the disk
temperature, which we fix to $T_1$
(Table~\ref{table:temperature}). Because our $T_1$ differs slightly from
EB94's $T_2, $ and also because {\sc diskbb} overestimates the
temperature by $\sim5$\% \citep{Zimmerman:05}, we test the effect on the
spin parameter of increasing or decreasing the disk temperature by a
factor of 2.  As the three rightmost columns in
Table~\ref{table:temperature} show, even such gross changes in disk
temperature have a negligible effect ($<0.5$\%) on the spin parameter.

\section{Comprehensive Error Analysis: Gold Spectra} 
\label{sect:error}

We now estimate the error in $a_*$ resulting from the uncertainties in
the input parameters $M$, $i$, $D$ and $N_{\rm H}$, which dominate the
error budget (including uncertainties in the model) in measuring spin
via the continuum-fitting method \citep[see ][ and references
therein]{McClintock:14}.  We first quantitatively explore for each
parameter separately the effect of its uncertainty on the fitted value
of the spin parameter.  Then, for the four gold spectra we describe our
standard Monte Carlo (MC) error analysis and present our adopted final
results.  We perform our MC analysis on each spectrum separately,
which is our usual approach, and we also fit the four gold spectra
jointly. 

\subsection{Effect of Varying the Input Parameters
  Individually}\label{sec:para}

In turn, we fixed three of the input parameters $M$, $i$, $D$ and
$N_{\rm H}$ and varied the fourth in order to assess its effect on the
best-fit value of $a_*$. The results, which are shown in Figure
\ref{fig:error1}, demonstrate that the value of the spin parameter is
most sensitive to uncertainties in the distance, followed in succession
by uncertainties in mass and inclination.  As expected, the uncertainty
in $N_{\rm H}$ is relatively unimportant because the column density is
modest and the detector is unresponsive below 2 keV
(Section~\ref{sect:data}).

\subsection{MC Error Analysis: Individual Fits}\label{sec:MC}

The MC method has long been our standard approach to error analysis
\citep{Liu:08,Gou:09,Gou:10}.  Here, we consider only the four gold
spectra and the effects of uncertainties in the four parameters $M$,
$i$, $D$ and $N_{\rm H}$. In our analysis, we assumed that $N_{\rm H}$
is Gaussian distributed, and we used the log-normal function to describe
the asymmetric distributions of the parameters $M$, $i$, and $D$.  We
first fixed the viscosity parameter to our baseline value,
$\alpha=0.1$. Following the prescription described in
\citet{Gou:09,Gou:10}, for each of the four spectra (SP1--SP4 in
Table~\ref{table:observations}) we (1) generated 3000 sets of the four
parameters assuming that they are independent and normally distributed;
(2) computed for each set of input parameters the {\sc kerrbb2} look-up
table for the spectral hardening factor $f$; and (3) fitted the spectrum
to determine $a_*$.

We then performed the MC analysis for $\alpha=0.01$ following precisely
the same procedures.  The resultant histograms for $\alpha=0.01$ and
$\alpha=0.1$ showing the number of occurrences vs.\ the spin parameter,
and a summation of the two histograms, are presented in Figure
\ref{fig:comparison} by a dashed line, a thin solid line and a thick
solid line, respectively.  Clearly, the effect of varying the viscosity
parameter is slight. Adopting the summed histogram, {\it we arrive at
  our final adopted estimate of the spin parameter:
  $a_*=0.63_{-0.19}^{+0.16}$} (1$\sigma$ level of
confidence). Uncertainty ranges at three other levels of confidence are
summarised in Table~\ref{table:spin}. The effect on the spin parameter
of varying individually the four input parameters is illustrated in
Figure~\ref{fig:error2}.

\subsection{MC Error Analysis: Joint Fit}\label{sec:joint}
 
We now show that the alternative approach of fitting the four gold
spectra simultaneously produces results that are essentially identical
to our adopted results (which were obtained by fitting the spectra
individually). In this case, we allowed all of the parameters to vary
freely except the parameter of interest, namely, the spin parameter,
which has a unique physical value. Repeating the MC analysis using
exactly the same procedures described above, we arrived at precisely the
same estimate of spin as before at the 1$\sigma$ level of confidence:
$a_*=0.63_{-0.19}^{+0.16}$. Meanwhile, as summarized in the rightmost
column of Table~\ref{table:spin}, the uncertainty ranges at other levels
of confidence differ very slightly from our adopted results.

\section{Comprehensive Error Analysis: Complete Sample of 15
  Spectra} \label{sect:error15}

In order to confirm our prime result for the four gold spectra, we
performed an MC error analysis on our complete sample of 15 spectra
employing exactly the same procedures described in
Section~\ref{sec:MC}. The spectra were analyzed individually. Their
histograms are shown in Figure~\ref{fig:MC_all}. The final result is the
summed histogram in Figure~\ref{fig:histo} plotted as a heavy solid
line, which is to be compared directly to its counterpart histogram in
Figure~\ref{fig:comparison}. In this case, the uncertainty in the spin
parameter is slightly less, but the central value is essentially
unchanged, which confirms our adopted result for the four gold spectra.
Figure~\ref{fig:error3} (like Figure~\ref{fig:error2} for the gold
spectra) shows the effect on the spin parameter of varying individually
the four input parameters.

\section{Discussion and Conclusions}  \label{sect:discussion}
 
Based on a sample of four sources, \citet[][~hereafter NM12]{Narayan:12}
proposed a relationship between jet power and black hole spin that is
consistent with the prediction of the B-Z jet model
\citep{Blandford:77}.  \citet{Steiner:13} confirmed the relation with a
fifth source and used the NM12 model to predict the spins for an
additional six black holes including \NM.  The expression for the
dimensionless jet power is:

\begin{equation}
  P_{\rm jet}=\left(\frac{\nu}{5~{\rm GHz}}\right)~\left(\frac{S_{\rm \nu,0}^{\rm tot}}{\rm Jy}\right)~\left(\frac{D}{\rm kpc}\right)^2~\left(\frac{M}{M_{\odot}}\right)^{-1},
\end{equation}
\noindent where $S_{\rm \nu,0}^{\rm tot}$ is the beaming-corrected flux,
\begin{equation}
S_{\rm \nu,0}^{\rm tot}=S_{\rm \nu,obs}\times\delta^{k-3},
\end{equation}

\noindent $\nu$ is the observing frequency, $D$ the distance, $M$ the
black hole mass, and $\delta$ the beaming correction factor. The
parameter $k$ is the radio spectral index whose value for NovaMus is 0.5
\citep{Ball:95}. For the approaching and receding jets,
$\delta=(\Gamma(1-\beta\cos{i}))^{-1}$ and
$\delta=(\Gamma(1+\beta\cos{i}))^{-1}$, respectively
~\citep{Mirabel:99}.

Figure \ref{fig:jet} is a revised version of Figure~1 from
\citet{Steiner:13} that shows jet power vs.\ spin for two typical values of
$\Gamma$.  Ignoring the point for NovaMus (red open square), the
difference between our figure and the one of Steiner et al.\ is that we
use for GRS~1915+105 new estimates of the key parameters:
$M=12.4~M_{\sun}, D=8.6$~kpc, $i=60$~deg and
$a_*=0.98_{-0.02}^{+0.01}$~\citep{reid:14}. Using these data and the
relations above, we re-estimated the jet power for GRS~1915+105 and
updated the corresponding data point in Figure \ref{fig:jet}. We then
re-fitted the data for the five sources (blue filled circles) using the
least $\chi^2$ method; the resultant relation describing the model
curves in Figure \ref{fig:jet} is:

\begin{displaymath}
P_{\rm jet} = \left(\frac{a_*}{1+\sqrt{1-{a_*}^2}}\right)^2~{\rm Jy}\\
~\times \left\{ \begin{array}{lll}
 {\rm Exp(3.9\pm0.5)} & \textrm{($\Gamma=2$)}\\
 {\rm Exp(6.9\pm0.5)} & \textrm{($\Gamma=5$)}.
 \end{array} \right.
\end{displaymath}

In estimating the jet power of NovaMus, for the 5 GHz flux we use values
that range from 0.2~Jy to 1.0~Jy \citep[see Footnote 1 of Table 1
in][]{Steiner:13}, which we take to be a 1$\sigma$ range.  As the
central value, we adopt the geometric mean of these two values, namely
0.45~Jy.  The flux is therefore $S_{\rm \nu,
  obs}=0.45_{-0.25}^{+0.55}$~Jy or, equivalently, log($S_{\rm \nu,
  obs}$/Jy)$=-0.35\pm0.35$. Using the equations above, the corresponding
values of jet power are $0.35_{-0.19}^{+0.43}$ for $\Gamma=2$ and
$3.5_{-1.9}^{+4.3}$ for $\Gamma=5$. Using these estimates of jet power
and our MC estimate of spin for NovaMus, we added a sixth data point to
Figure \ref{fig:jet}. Considering the uncertainties in both the model
and the data, the point for NovaMus lies off the model curves by
1.9$\sigma$ for $\Gamma=2$ and 2.1$\sigma$ for $\Gamma=5$.

\citet[][~M14]{Morningstar:14} reported a retrograde spin for the black
hole in \NM: $a_*=-0.25^{+0.05}_{-0.64}$ (90\% confidence level). As
discussed in detail by \citet{Fragos:15}, this is a surprising result.
In contrast, we find a moderately high value of spin, $a_* =
0.63_{-0.19}^{+0.16}$, and rule strongly against a retrograde value:
$a_*> 0.17$ ($2\sigma$ or $95.4$\% confidence level).

We and M14 analyzed the same {\it Ginga} X-ray data and we both used the
continuum-fitting method. Ignoring the crudeness of M14's error analysis
and treatment of spectral hardening, the crucial difference between our
study and theirs is the choice of the key input parameters: The values
they gleaned from the literature were $M=7.24\pm0.07~M_{\sun}$, $i=54
\pm1.5$ deg, and $D=5.89\pm0.26$ kpc. Meantime, our adopted values are
$M=11.0^{+2.1}_{-1.4}\p M_\sun$,\p$\rm i=43.2^{+2.1}_{-2.7}\p\rm
deg$,\p$ D=4.95^{+0.69}_{-0.65}\p\rm kpc$ \citep{Wu:15b}.  As
Figure~\ref{fig:error1} makes clear, each of M14's input parameters {\it
  considered individually} -- specifically, their 50\% lower mass, 10.8
deg higher inclination and 16\% greater distance -- is responsible for
driving their spin estimate downward, and for their conclusion that the
spin is retrograde.  The principal reference they cite for the values of
$M$, $i$ and $D$ they adopt is a one-page conference paper \citep{Gelino:04},
 which in turn is based on \citet{Gelino:01}.  This latter
paper, which arrives at an inflated estimate of the inclination, and
hence low value for the mass, is based on a flawed analysis of the
ellipsoidal light curves that ignores a substantial contribution of
light from the accretion disk \citep[see Sections 3 and 5 in ][]{Wu:15a}.

The X-ray data we have used in estimating the spin of NovaMus via the
continuum-fitting method are of extraordinary quality. In particular,
despite the excellent performance of the {\it Ginga} LAC detectors at
high energies, the four gold spectra show essentially no evidence of a
power-law component.  These pure thermal spectra, which are ideal for
application of the continuum-fitting method, our accurate estimates of
$M$, $i$ and $D$, and the demonstrated robustness of the
continuum-fitting method provide a firm estimate of the spin of the
black hole in \NM: $a_* = 0.63_{-0.19}^{+0.16}$.  This result is
confirmed by our analysis of eleven additional spectra of lower quality.



\begin{deluxetable}{cccccc}
\tablecolumns{6} 
\tablecaption{{\it Ginga} X-ray Observations of Nova Muscae 1991}
\tabletypesize{\small}
\tablehead {
\colhead{N}&\colhead{Time}&\colhead{Exposure Time}&\colhead{Counts}& \colhead{$L/L_{\rm Edd}$}\\
 &\colhead{(1991 UT)} &\colhead{(seconds)}
      }
\startdata
1&$2/13\p\p 5:06-\p 5:23$&391& 10265400&0.201\\
2&$2/13\p\p 6:40-\p 6:57$ &412& 10236100&0.197\\
3&$2/13\p\p 8:14-\p 8:34$ &598& 14826300&0.196\\
4&$2/14\p\p 4:36-\p 4:53$ &219&  5208070&0.189\\
5&$2/14\p\p 5:30-\p 5:34$ & 31&  739773&0.191\\
6&$2/20\p23:31-23:37$   & 83&  1619840&0.163\\
7&$2/21\p\p 0:23-\p 0:30$ &166&  3212070&0.162\\
8&$3/08\p18:04-18:22$ &472&  5285530&0.103\\
9&$3/10\p16:56-17:17$   &321&  3303930&0.093\\
10&$3/20\p12:56-13:17$   &532&  5175480&0.090\\
11&$3/28\p\p 9:37-\p 9:43$ &140&  1607220&0.103\\
12&$3/29\p\p 5:54-\p 6:06$&457&  5251350&0.104\\
13&$3/30\p\p 8:36-\p 8:54$ &655&  7187840&0.099\\
14&$4/02\p\p5:04-\p5:30$&934& 10122400&0.099\\
15&$4/19\p21:06-21:20$  & 31&  179975&0.064\\
16&$5/17\p\p 3:12-\p 3:20$ &291&  664975&0.031\\
17&$5/17\p\p 4:34-\p 4:57$ &681&  1263360&0.010\\
18&$5/17\p\p 7:49-\p 8:10$ &978&  2387950&0.033\\
19&$5/18\p\p 2:18-\p 2:40$ &691&  1415900&0.018\\
20&$5/18\p\p 3:52-\p 4:17$ &1144& 1776410&0.010\\
21&$5/18\p\p 5:25-\p 5:52$ &1330& 3368750&0.031
 \enddata
 
\tablecomments{All the spectra analyzed by EB94 with Eddington-scaled
  luminosities between 1\% and 30\% are listed here. In our analysis, we
  ignore six of these spectra (Nos. 5, 15, 16, 16, 19 and 20) for the
  reasons given in the text.}
\label{table:time}
\end{deluxetable}

\begin{deluxetable}{ccccccccc}
\tablecolumns{9} 
\tablecaption{Fit Results for Nova Muscae 1991}
\tabletypesize{\tiny}
\tablehead {
\colhead{}&\colhead{N}&\multicolumn{2}{c}{SIMPLR}&\multicolumn{2}{c}{KERRBB2}& \colhead{$\chi_{\rm \nu}^2/{\rm d.o.f.}$}& \colhead{$f$}& \colhead{$L/L_{\rm Edd}$}\\
 &\colhead{} & \colhead{$\Gamma$} & \colhead{$f_{\rm SC}$} & \colhead{$a_*$} & \colhead{$\dot{M}$} 
      }
 \startdata
 
SP1&11&$ 1.63\pm 0.46$&$0.00034\pm0.00024$&$ 0.62\pm 0.01$&$ 1.70\pm 0.03$&$1.70/23$&1.63&0.103\\
SP2&12&$ 1.90\pm 0.67$&$0.00022\pm0.00024$&$ 0.61\pm 0.01$&$ 1.74\pm 0.03$&$0.98/24$&1.63&0.104\\
SP3&13&$ 1.63\pm 0.47$&$0.00016\pm0.00012$&$ 0.64\pm 0.01$&$ 1.60\pm 0.03$&$0.82/27$&1.63&0.099\\
SP4&14&$ 1.96\pm 0.30$&$0.00041\pm0.00020$&$ 0.62\pm 0.01$&$ 1.65\pm 0.03$&$1.07/28$&1.63&0.099\\

\tableline
SP5 & 1& 2.24 $\pm$ 0.05&0.057 $\pm$ 0.0048&0.72 $\pm$ 0.02 &2.30 $\pm$ 0.08&1.14/27 &1.68&0.201 \\ 
SP6 & 2& 2.11 $\pm$ 0.05&0.034 $\pm$ 0.0030&0.63 $\pm$ 0.02 &2.66 $\pm$ 0.08&0.81/27 &1.67&0.198 \\ 
SP7 & 3& 2.00 $\pm$ 0.04&0.026 $\pm$ 0.0022&0.65 $\pm$ 0.02 &2.62 $\pm$ 0.09&1.72/27 &1.67&0.198 \\ 
SP8 & 4& 2.38 $\pm$ 0.06&0.034 $\pm$ 0.0038&0.47 $\pm$ 0.03 &3.32 $\pm$ 0.13&1.62/27 &1.67&0.194 \\ 
SP9 & 6& 2.23 $\pm$ 0.09&0.022 $\pm$ 0.0043&0.54 $\pm$ 0.03 &2.70 $\pm$ 0.12&0.45/27 &1.66&0.166 \\ 
SP10& 7& 2.39 $\pm$ 0.05&0.032 $\pm$ 0.0028&0.53 $\pm$ 0.02 &2.73 $\pm$ 0.10&0.66/27 &1.66&0.165 \\ 
SP11& 8& 2.37 $\pm$ 0.38&0.004 $\pm$ 0.0037&0.55 $\pm$ 0.02 &1.86 $\pm$ 0.07&0.99/27 &1.63&0.104 \\ 
SP12& 9& 2.71 $\pm$ 0.54&0.004 $\pm$ 0.0055&0.53 $\pm$ 0.02 &1.73 $\pm$ 0.07&0.73/24 &1.62&0.093 \\ 
SP13&10& 2.94 $\pm$ 0.68&0.005 $\pm$ 0.0073&0.61 $\pm$ 0.01 &1.53 $\pm$ 0.04&0.94/25 &1.62&0.090 \\ 
SP14&18& 2.35 $\pm$ 0.05&0.033 $\pm$ 0.0040&0.54 $\pm$ 0.02 &0.67 $\pm$ 0.03&1.02/27 &1.52&0.033 \\ 
SP15&21& 2.35 $\pm$ 0.04&0.066 $\pm$ 0.0063&0.62 $\pm$ 0.02 &0.55 $\pm$ 0.03&0.81/27 &1.52&0.032 

 \enddata
\normalsize
\tablecomments{
From left to right, and for the model components indicated, the
columns contain: (1) The name of the spectrum; (2) the corresponding
index number from Table~1; (3) photon power-law index; (4) scattering
fraction; (5) spin parameter; (6) mass accretion rate in units of
$10^{18}$~g~s$^{-1}$; (7) reduced chi-square and degrees of freedom; (8)
spectral hardening factor; (9) bolometric Eddington-scaled X-ray
luminosity of the thermal disk component.  For the silver spectra
SP5--SP15 (which include a reflection component modeled using {\sc
  ireflect} and {\sc kerrdisk}), the values of three additional fit
parameters are given in Table~\ref{table:observations_2}.  
}

\label{table:observations}
\end{deluxetable}


\begin{deluxetable}{cccccccc}
\tablecolumns{8} 
\tablecaption{Influence of Disk Temperature on the Spin Parameter}
\tabletypesize{\small}

\tablehead {
\colhead{}&\colhead{N}&\multicolumn{3}{c}{T(KeV)}&\multicolumn{3}{c}{spin}\\
&\colhead{} & \colhead{$T_1$}&\colhead{$T_2$}&\colhead{relative difference}&\colhead{$a_*(T_1)$}&\colhead{$a_*(2T_1)$}&\colhead{$a_*(0.5T_1)$}
}
 \startdata
SP1&11&0.713   &0.716&0.4\%&$-$&$-$&$-$\\
SP2&12&0.719 &0.720&0.1\%&$-$&$-$&$-$\\
SP3&13&0.727   &0.733&0.8\%&$-$&$-$&$-$\\
SP4&14&0.720&0.716&0.5\%&$-$&$-$&$-$\\
SP5&1&$0.831$&0.810&2.6\%& 0.7134& 0.7143& 0.7130\\
SP6&2&$0.800$&0.784&2.0\%& 0.6338& 0.6340& 0.6311\\
SP7 &3&$0.844$&0.795&6.2\%& 0.6447& 0.6443& 0.6450\\
SP8 &4&0.777&0.757&2.6\%& 0.5131& 0.5144& 0.5126\\
SP9 &6&$0.762$  &0.740&3.0\%& 0.5735& 0.5735& 0.5729\\
SP10&7&$0.760$&0.735&3.4\%& 0.5422& 0.5438& 0.5416\\
SP11&8&$0.702$&0.707&0.7\%& 0.5634& 0.5633& 0.5634\\
SP12&9&$0.674$  &0.678&0.5\%& 0.5434& 0.5434& 0.5435\\
SP13&10&$0.674$ &0.702&4.0\%& 0.6166& 0.6165& 0.6169\\
SP14&18&$0.506$&0.502&0.8\%& 0.5428& 0.5445& 0.5426\\
SP15&21&$0.510$&0.497&2.6\%& 0.6282& 0.6302& 0.6278

 \enddata
 \tablecomments{$T_1$ is our fitted value of temperature and $T_2$ is
   the value reported by EB94; N is the index number given in Table 1.
 }
\label{table:temperature}
\end{deluxetable}

\begin{deluxetable}{ccccccccc}
\tablecolumns{9} 
\tablecaption{Fit Results for the Reflection Components for the Silver Spectra}
\tabletypesize{\tiny}
\tablehead {
\colhead{}&\colhead{N}&\multicolumn{2}{c}{KERRDISK}&\multicolumn{2}{c}{IREFLECT}& \colhead{$\chi_{\rm \nu}^2/{\rm d.o.f.}$}& \colhead{$f$}& \colhead{$L/L_{\rm Edd}$}\\
 &\colhead{} & \colhead{$E_{\rm L}$}& \colhead{$N_{\rm L}$} & \colhead{$T$} & \colhead{$\xi$}  
      }
 \startdata

SP5 & 1& 6.97$\pm$ 0.31&0.038 $\pm$ 0.008&0.83 &100000 $\pm$ 249879&1.14/27 &1.68&0.201 \\ 
SP6 & 2& 6.97$\pm$ 0.25&0.029 $\pm$ 0.006&0.81 &100000 $\pm$ 264214&0.81/27 &1.67&0.198 \\ 
SP7 & 3& 6.97$\pm$ 0.19&0.026 $\pm$ 0.007&0.83 &100000 $\pm$ 275646&1.72/27 &1.67&0.198 \\ 
SP8 & 4& 6.46$\pm$ 0.35&0.009 $\pm$ 0.005&0.78 &   279 $\pm$    276&1.62/27 &1.67&0.194 \\ 
SP9 & 6& 6.66$\pm$ 0.70&0.004 $\pm$ 0.004&0.76 &  1598 $\pm$   1567&0.45/27 &1.66&0.166 \\ 
SP10& 7& 6.50$\pm$ 0.39&0.006 $\pm$ 0.004&0.76 &   184 $\pm$    135&0.66/27 &1.66&0.165 \\ 
SP11& 8& 6.62$\pm$ 0.31&0.002 $\pm$ 0.001&0.70 &   749 $\pm$   5442&0.99/27 &1.63&0.104 \\ 
SP12& 9& 6.40$\pm$ 1.16&0.001 $\pm$ 0.001&0.67 &  1077 $\pm$  13609&0.73/24 &1.62&0.093 \\ 
SP13&10& 6.97$\pm$ 0.58&0.001 $\pm$ 0.001&0.69 &   872 $\pm$  12932&0.94/25 &1.62&0.090 \\ 
SP14&18& 6.55$\pm$ 0.26&0.001 $\pm$0.0004&0.51 &   343 $\pm$    217&1.02/27 &1.52&0.033 \\ 
SP15&21& 6.44$\pm$ 0.33&0.002 $\pm$ 0.001&0.51 &   390 $\pm$    227&0.81/27 &1.52&0.032

 \enddata
\normalsize

\tablecomments{
From left to right, and for the model components indicated, the columns
contain: (1) The name of the spectrum; (2) the corresponding index
number from Table~1; (3) the central line energy in keV limited from 6.4
to 6.97; (4) line flux in units of photons~cm$^{-2}$~s$^{-1}$; (5) disk
temperature in keV; (6) ionization parameter in units of
erg~cm$^{-2}$~s$^{-1}$; (7) reduced chi-square and degrees of freedom;
(8) spectral hardening factor; (9) bolometric Eddington-scaled X-ray
luminosity of the thermal disk component. For all spectra, we adopt the
standard value of the emissivity index, $q=3$. For spectra SP5--7, the
disk ionization parameter is pegged at its maximum value,
$\xi$=10$^5$~erg~cm$^{-2}$~s$^{-1}$.  
}
\label{table:observations_2}
\end{deluxetable}

\begin{deluxetable}{lcc}
\tablecolumns{2} 
\tablecaption{Spin Determinations for Different Confidence Levels}
\tabletypesize{\small
}
\tablehead {
\colhead{Confidence Level}&\multicolumn{2}{c}{Spin Interval~($a_*$)}\\
&\colhead{MC Method (Adopted)}&\colhead{MC Method (Joint Fit)}
		}
 \startdata
 $68.3\%(1\sigma)$	& $0.63_{-0.19}^{+0.16}$&$0.63_{-0.19}^{+0.16}$\\
 90\%				& $0.63_{-0.25}^{+0.19}$&$0.63_{-0.24}^{+0.20}$\\
 $95.4\%(2\sigma)$	& $0.63_{-0.42}^{+0.26}$&$0.63_{-0.40}^{+0.26}$\\
 $99.7\%(3\sigma)$	& $0.63_{-0.70}^{+0.34}$&$0.63_{-0.68}^{+0.33}$
 \enddata
 \tablecomments{
Confidence levels for our four gold spectra SP1--SP4 resulting from our
MC error analysis for both the fits to the individual spectra and for
the joint fit. The results are marginalized over our two fiducial values
of the viscosity parameter, $\alpha=0.1$ and $\alpha=0.01$.
}
\label{table:spin}
\end{deluxetable}

\begin{figure*}[f!]
    \begin{center}
      \includegraphics[angle = 0, trim =0cm 0cm 0cm 0cm,width=1.0\textwidth]{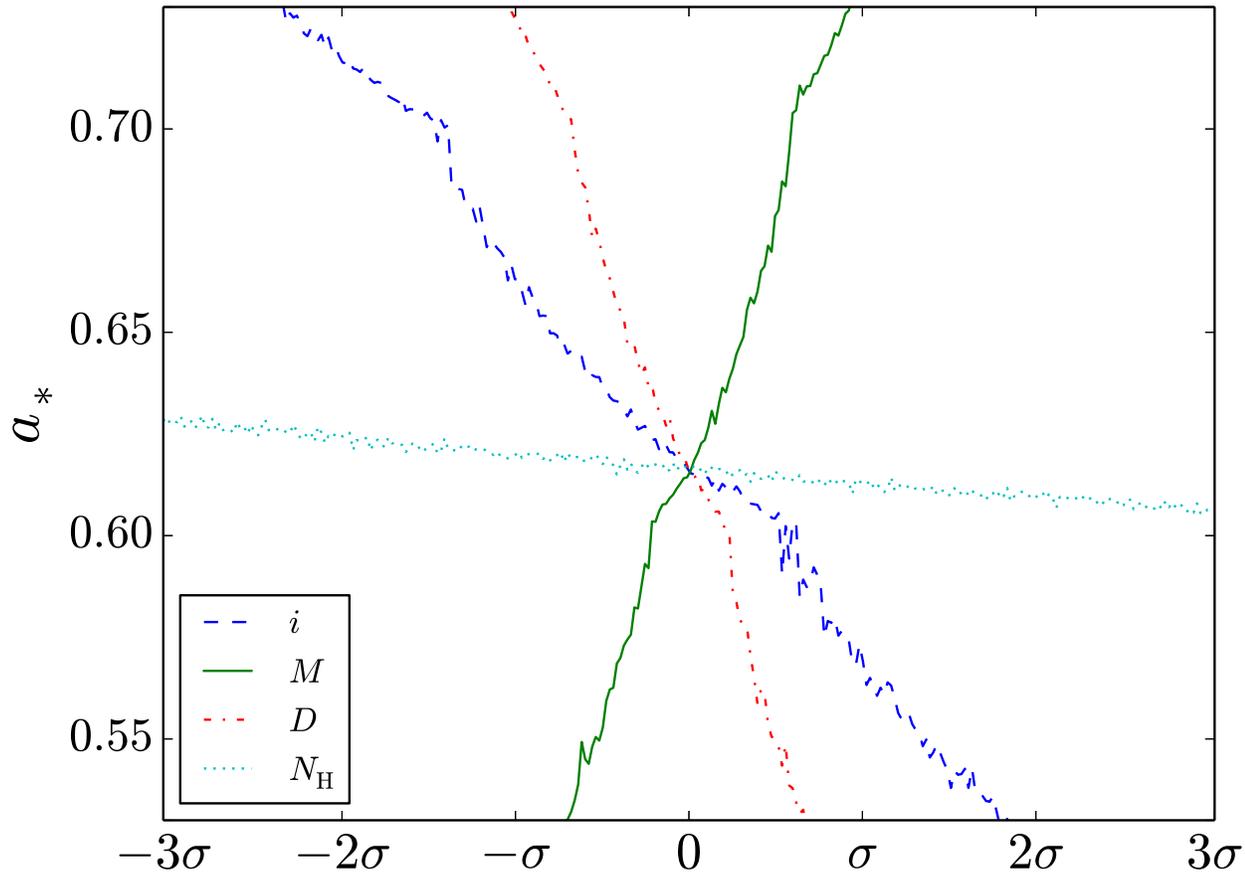}
       \end{center}
\caption{
Dependence of the spin parameter $a_*$ on each of the four input
parameters considered individually. The difference between the value of
a given parameter and its fiducial value is expressed in standard
deviations. The jaggedness of the lines is an artifact of the MC
sampling.
}
\label{fig:error1}
\end{figure*}

\begin{figure*}[f!]
    \begin{center}
      \includegraphics[angle = 0, trim =0cm 0cm 0cm 0cm,width=1.0\textwidth]{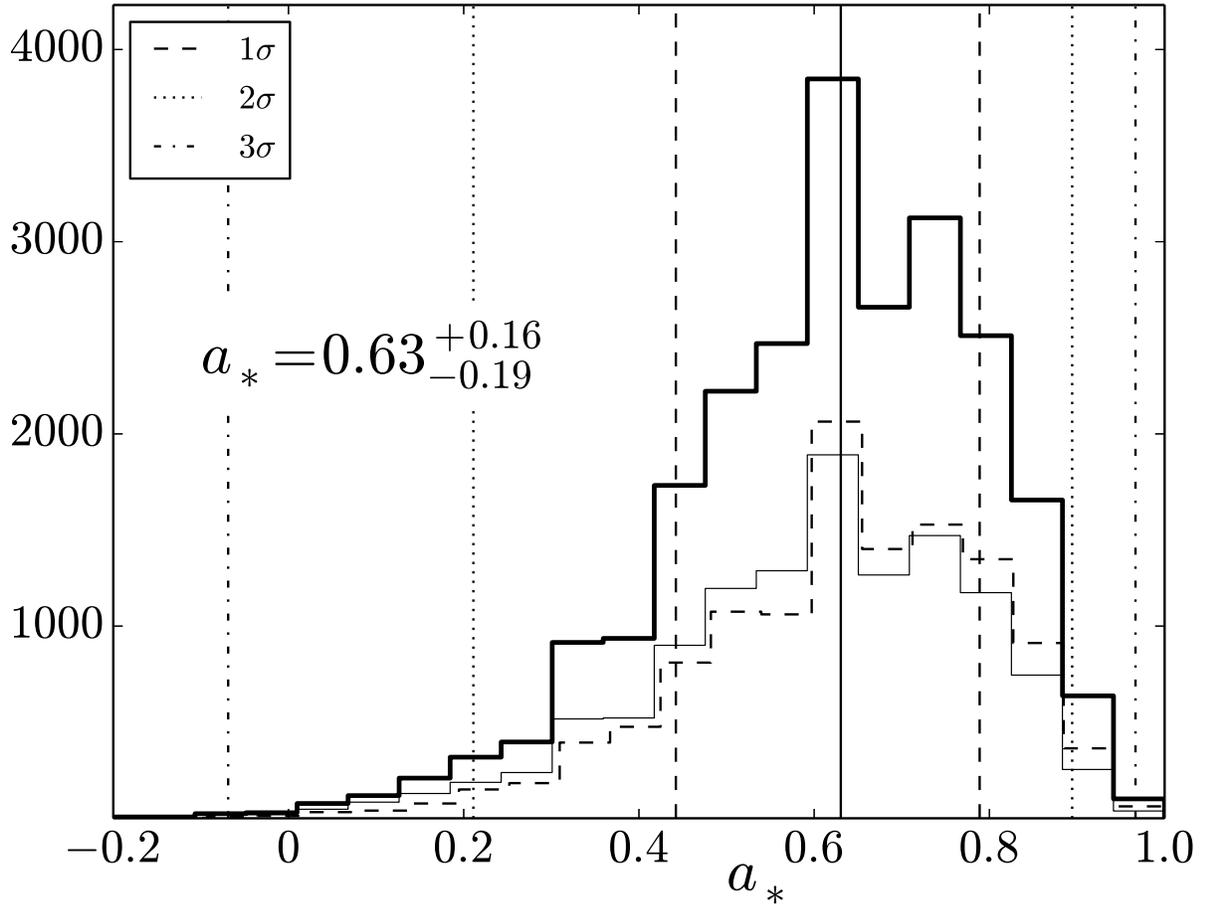}
       \end{center}
\caption{
Histograms of the distribution of the spin parameter computed using the
MC method for our gold spectra SP1--SP4
(Table~\ref{table:observations}). The histograms plotted using a thin
solid line and a dashed line were computed for $\alpha$ = 0.1 and
$\alpha$= 0.01, respectively. The summation of these two histograms is
shown as a bold solid line. The histograms were computed for 3000
parameter sets for each spectrum and for each of the values of the
viscosity parameter, for a total of 24,000 data points.  The estimate of
the spin parameter given is our final, adopted value.
}
\label{fig:comparison}
\end{figure*}

\begin{figure*}[f!]
   \begin{center}
     \includegraphics[angle = 0, trim =0cm 0cm 0cm 0cm,width=1.0\textwidth]{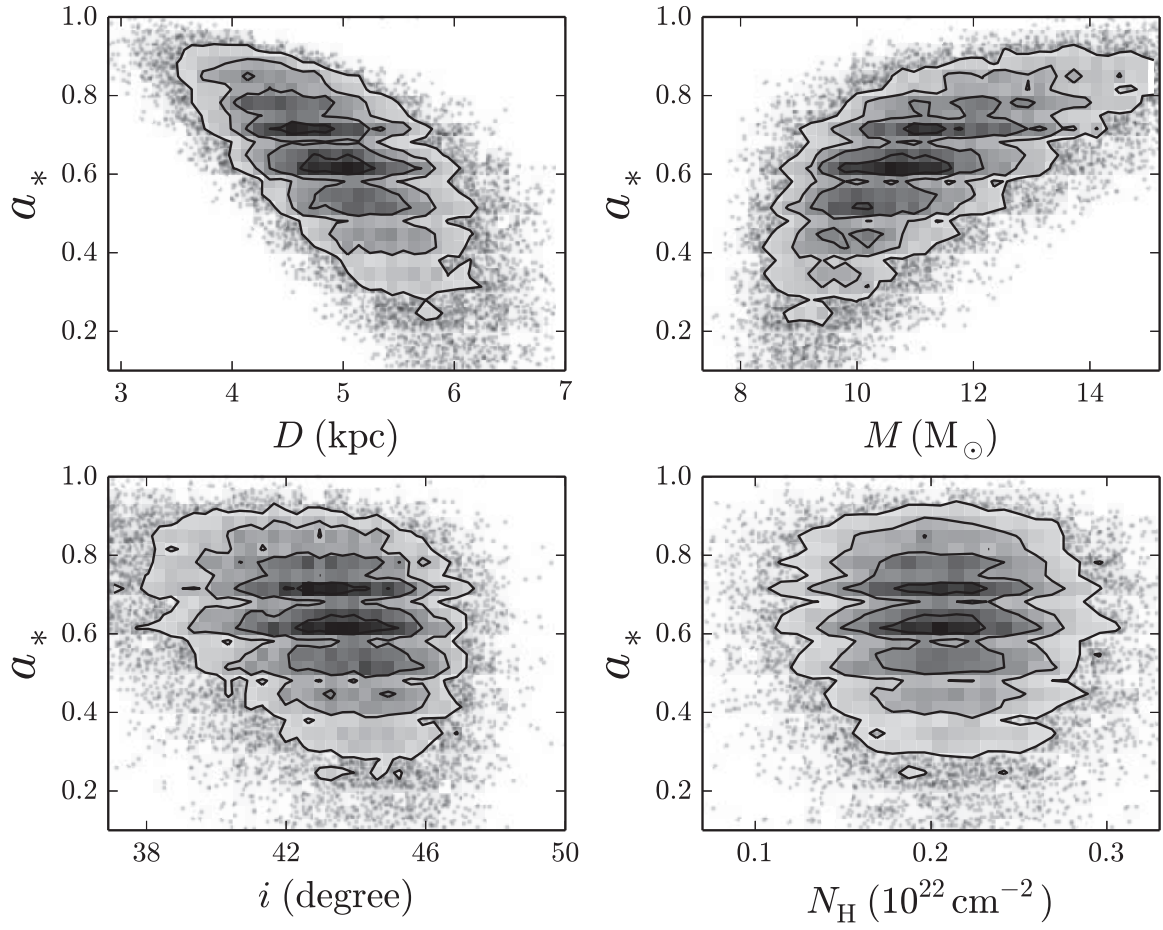}
      \end{center}
\caption{   
  Correlation plots for the MC method showing the effect
  on the spin parameter of varying $M, i, D$ and $N_{\rm H}$, while
  marginalizing over our two fiducial values of the viscosity
  parameter. Each panel contains 24,000 data points (see
  Figure~\ref{fig:comparison}).}
\label{fig:error2}
\end{figure*}

\begin{figure}[f!]
\epsscale{1} 
\plotone{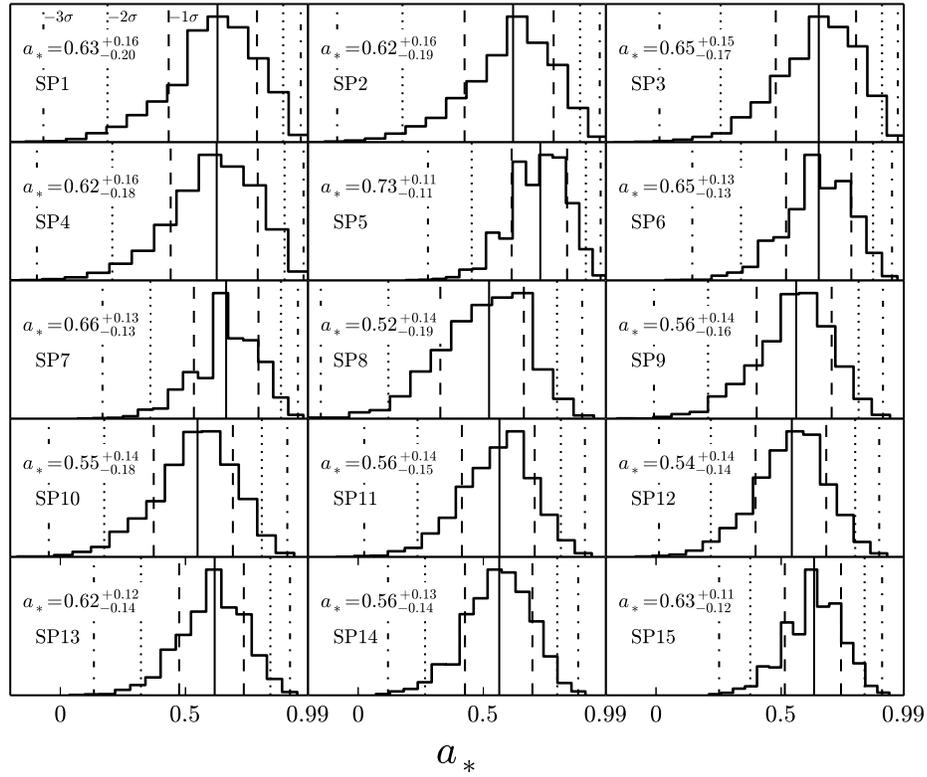}
\caption{ 
Results of our MC error analysis for the complete sample of 15
  spectra. The histogram in each panel is a summation of a pair of
  histograms, one computed for $\alpha=0.1$ and the other for
  $\alpha=0.01$ (see Figure~\ref{fig:comparison}).  
}
\label{fig:MC_all}
\end{figure}

\begin{figure*}[f!]
    \begin{center}
      \includegraphics[angle = 0, trim =0cm 0cm 0cm 0cm,width=1.0\textwidth]{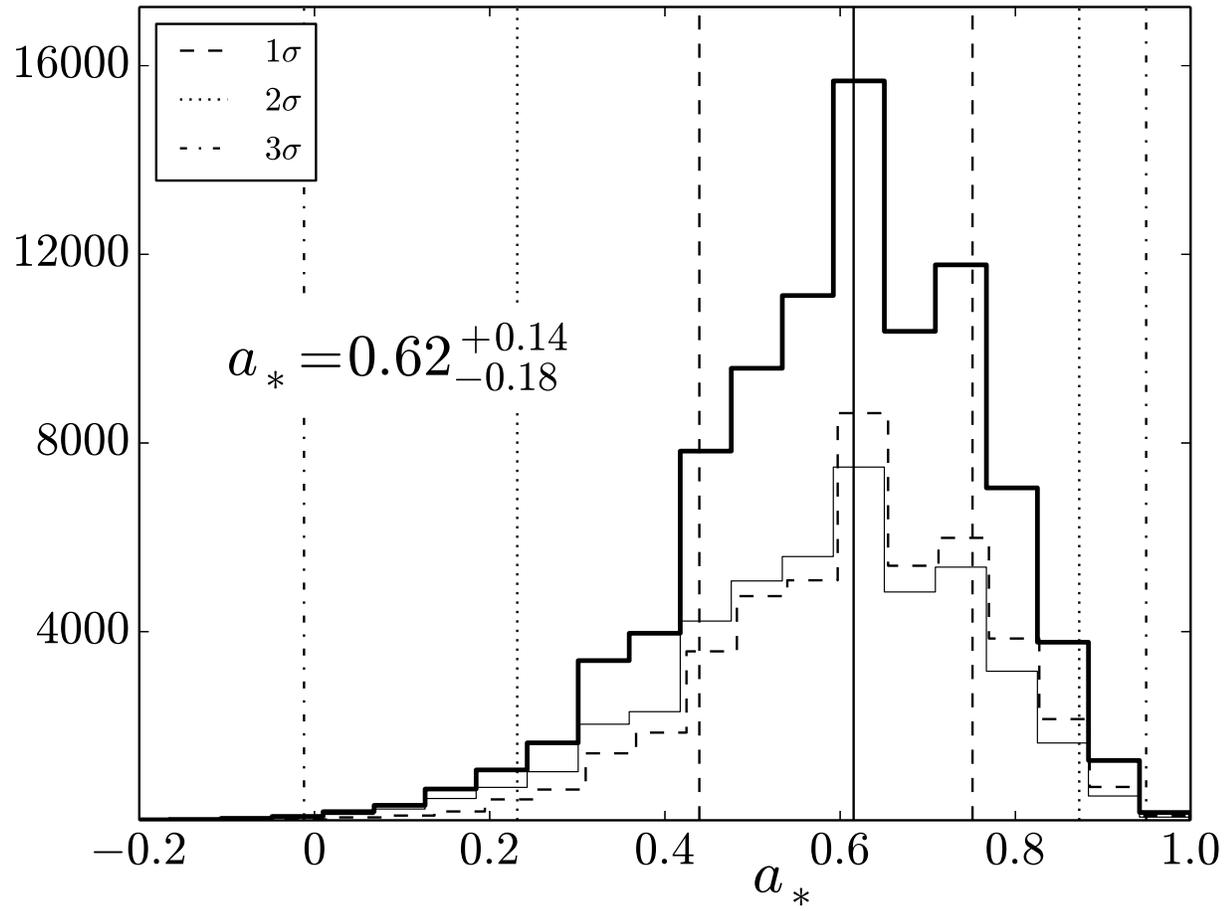}
       \end{center}
\caption{
Histograms of the distribution of spins generated by our MC error
analysis for the complete sample of 15 spectra. For a description of
this figure, see Figure~\ref{fig:comparison}.
}
\label{fig:histo}
\end{figure*}

\clearpage 
\begin{figure}[ht]
\epsscale{1} 
\plotone{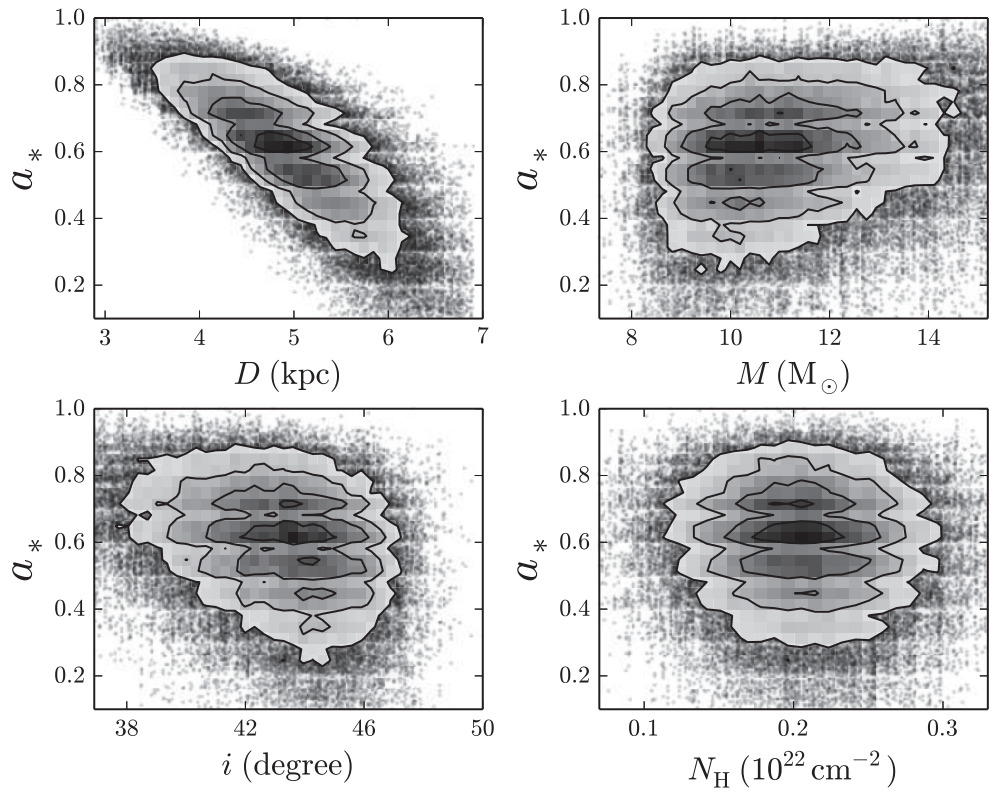}
\caption{Same as Figure~\ref{fig:error2} for the four gold
spectra except that these correlation plots result from the analysis
of our complete sample of 15 spectra. In this case, each panel
contains a total of 90,000 data points.}
\label{fig:error3}
\end{figure}

\begin{figure*}[f!]
    \begin{center}
      \includegraphics[angle = 0, trim =0cm 0cm 0cm 0cm,width=1.0\textwidth]{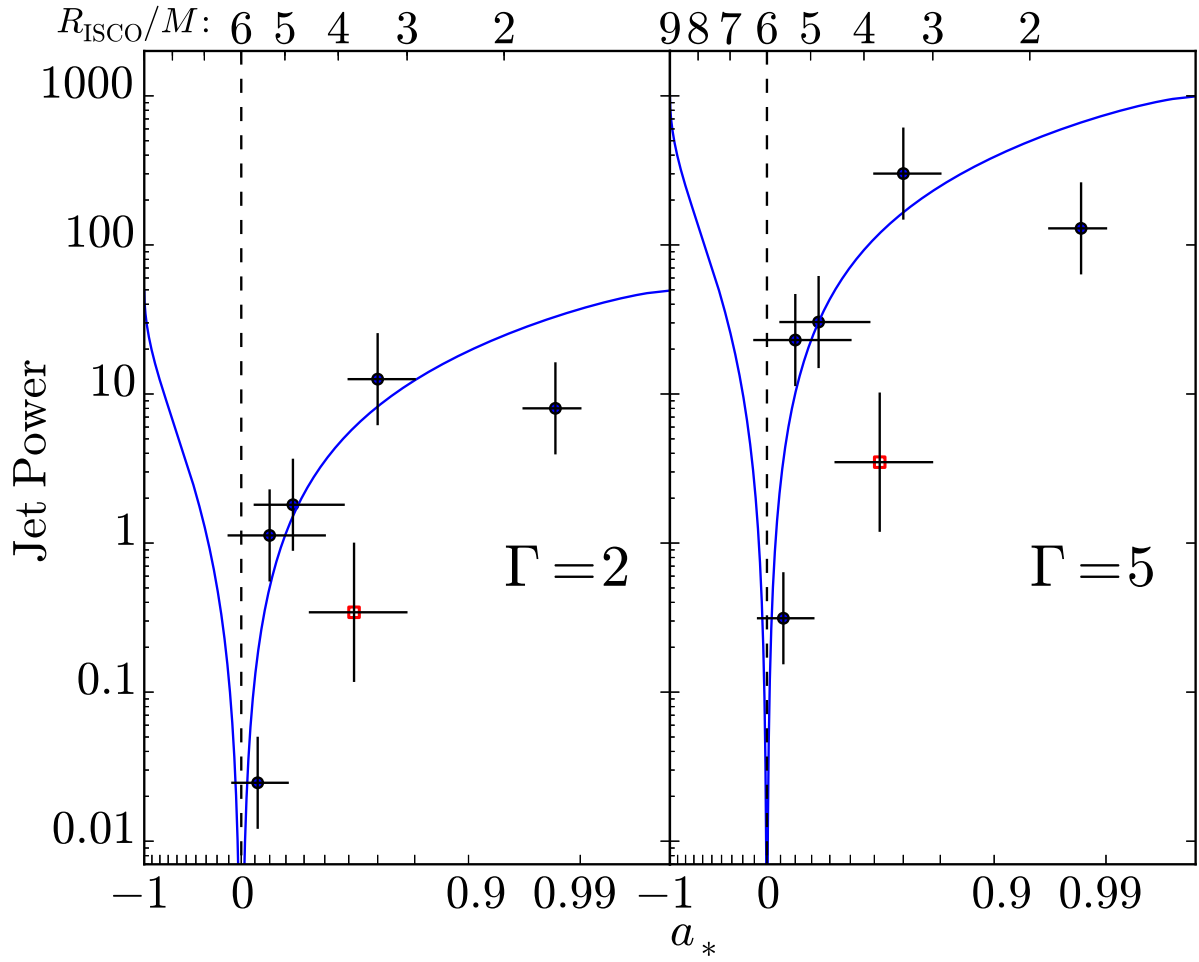}
       \end{center}
 \caption{Relationship between radio jet power and the
observable $R_{\rm ISCO} / M$ (top axis) and black hole spin (bottom
axis). The curve is fitted to the five data points plotted as solid
filled circles. The data point for NovaMus is plotted as an open red
square.}
\label{fig:jet}
\end{figure*}

\newpage
\clearpage

\end{document}